\documentstyle[11pt,moriond,epsfig]{article}

\def\Journal#1#2#3#4{{#1} {\bf #2}, #3 (#4)}

\def\be{\begin{equation}}
\def\ee{\end{equation}}
\def\bea{\begin{eqnarray}}
\def\eea{\end{eqnarray}}

\begin{document}
\vspace*{4cm}
\title{A GENERALIZED INFLATION MODEL\\
 WITH COSMIC GRAVITATIONAL WAVES}

\author{ V.N. LUKASH$^1$, E.V. MIKHEEVA$^1$, V. M\"ULLER$^2$, A.M. MALINOVSKY$^1$ }

\address{
$^1$ Astro Space Center of Lebedev Physical Institute,\\
      84/32 Profsoyuznaya, Moscow 117810, Russia\\
$^2$ Astrophysikalisches Institute Potsdam, An der Sternwarte 16, \\
     Potsdam, D-14482 Germany}

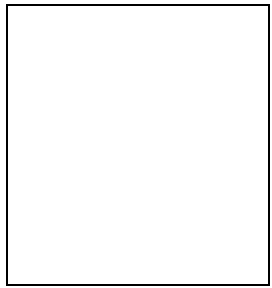
\maketitle\abstracts{We propose a $\Lambda$-inflation model which explains a 
large fraction of the COBE signal by cosmic gravitational waves. The primordial 
density perturbations fulfil both the contraints of large-scale microwave 
background and galaxy cluster normalization. The model is tested against the 
galaxy cluster power spectrum and the high-multipole angular CMB anisotropy.}

\section{Introduction}

The observational reconstruction of the {\it cosmological density 
perturbation} (CDP) spectrum is a key problem of the modern cosmology. It
provids a dramatic challenge after detecting the primordial CMB anisotropy as 
the signal found by DMR COBE at $10^0$ has appeared to be few times higher 
than the expected value of $\Delta T/T$ in the most simple and best developed 
cosmological standard CDM model.  

During recent years there were many proposals to improve sCDM (in the simplest
term, to remove the discrepancy between the CDP amplitudes at 
$8h^{-1} \rm{Mpc}$ as determined by galaxy clusters, and at large scales, 
$\sim 1000h^{-1} \rm{Mpc}$, according to $\Delta T/T$) by adding hot dark matter, a 
$\Lambda$-term, or considering non-flat primordial CDP spectra. Below, we 
present another, presumably more natural way to solve the sCDM problem based 
on taking into account a possible contribution of {\it cosmic gravitational 
waves} (CGWs) into the large-scale CMB anisotropy; we will also try to 
preserve the original near-scale-invariant CDP spectrum. Thus, the problem is 
reduced to the construction of a simple inflation producing near {\it 
Harrison-Zel'dovich} (HZ) spectrum of CDPs ($n_S\simeq 1$) and a considerable 
contribution of CGWs into the large-scale $\Delta T/T$.

A simple model of such kind is $\Lambda$-inflation, an inflationary model with
an effective metastable $\Lambda$-term\cite{lm1}${}^,$ \cite{lm3}. This model 
produces both S (CDPs) and T (CGWs) modes which have a non-power-law spectra, 
with a shallow minimum in the CDP spectrum, located at a scale $k_{cr}$ (there the 
$\Lambda$-term and the scalar field have equal energies while slowly-rolling at
inflation) where the S-slope
is exact HZ locally. The S-spectrum is 'red' for $k<k_{cr}$, and 'blue' for $k>k_{cr}$; around the $k_{cr}$ scale T/S is close to its maximum, 
it is of the order unity depending on the model parameters.

\section{$\Lambda$-inflation with self-interaction}

Let us consider a general potential of $\Lambda$-inflation driven by a single
scalar field $\varphi$:
\begin{equation}
V(\varphi) = V_0 + \sum_{\kappa=2}^{\kappa_{max}} \frac{\lambda_{\kappa}}
{\kappa}\varphi^{\kappa},
\end{equation}
where $V_0 > 0$ and $\lambda_{\kappa}$  are constants, $\kappa=2, 3, 4,..$.
In the case of massive inflaton ($\kappa = \kappa_{max} = 2$) T/S can be 
larger than unity only when the CDP spectrum 
slope in the `blue' asymptote is very steep, $n_S^{blue} > 1.8$. To avoid such
a strong spectral bend on short scales ($k > k_{cr}$) we choose here another 
simple version of $\Lambda$-inflation -- the case with self-interaction:  
$\kappa = \kappa_{max} = 4$, $\lambda_4 \equiv \lambda > 0$; this model is 
called $\Lambda\lambda$-inflation.

The scalar field $\varphi$ drives an inflationary evolution if $\gamma \equiv
- \dot H/H^2 < 1$, where $H \simeq \sqrt{V/3}$ (we assume  
$8\pi G = c =\hbar = 1$ and $H_0=100h\rm{km}\;\rm{s}^{-1}\rm{Mpc}^{-1}$). 
This condition holds true for all values of $\varphi$ if
\begin{equation} 
c \equiv \frac 14 \varphi_{cr}^2 =\frac 12
\sqrt{\frac{V_0}{\lambda}} > 1, 
\end{equation} 
which we imply hereafter. The fundamental gravitation perturbation spectra 
$q_k$ and $h_k$ generated in $\Lambda\lambda$-inflation in S and T modes, 
respectively, are as follows \cite{l1}${}^,$ 
\cite{l2}${}^,$ \cite{lm3}:
\begin{equation} 
q_k = \frac H{2\pi\sqrt{2\gamma}} = \frac{\sqrt{2\lambda/3}}{\pi}
\left(c^2+x^2\right)^{3/4},\;\;\;\;
h_k = \frac{H}{\pi\sqrt{2}} = \frac{2c\sqrt{\lambda/3}}{\pi} 
\left ( 1 + \frac x{\sqrt{c^2 + x^2}} \right)^{-1/2}, 
\end{equation}  
where 
\begin{equation}
x =
\ln \left[ \frac{k}{k_{cr}} \left(1+\left(\frac xc\right)^2\right)^{1/4} \left(
1+\frac x{\sqrt{c^2+x^2}}\right)^{2/3}\right]\simeq\ln(k/k_{cr}).
\end{equation}
The dimentionless spectrum of density perturbations depends on a transfer 
function $T(k)$:
\begin{equation}
\Delta_k =3.6\times 10^6\left(\frac{k}{h}\right)^2 q_k T(k).
\end{equation}
\section{CDM cosmology from $\Lambda\lambda$-inflation}

Let us consider the CDP spectrum with CDM transfer function, normalized both 
by $\Delta T/T\vert_{10^0}$ (including the contribution from 
CGWs) and the galaxy cluster abundance at $z = 0$, to find the family of the 
most realistic $q$-spectra produced in $\Lambda\lambda$-inflation.

In total, we have three parameters entering the function $q_k$ : $\lambda$, $c$
and $k_{cr}$. Constraining them by two observational tests we are actually 
left with only one free parameter (say, $k_{cr}$) which may be restricted 
elsewhere by other observations.

To demonstrate explicitly how the three parameters are mutually related, we 
first employ a simple analytical estimates for the $\sigma_8$ and $\Delta T/T$ 
tests to derive the key equation relating $c$ and $k_{cr}$, and then solve 
it explicitly to obtain the range of interesting physical parameters.

Instead of taking the $\sigma$-integral numerically we may estimate the spectrum 
amplitude on cluster scale ($k=k_1\simeq 0.3h/\rm{Mpc}$):
\begin{equation}
q_{k_1} \simeq 4.5\times 10^{-7}\frac{h^2 \sigma_8}{k_1^2 T(k_1)}\;.
\end{equation}

On the other hand, the spectrum amplitude on large scale ($k_2=k_{COBE}\simeq
10^{-3}h/\rm{Mpc}$) can be taken from $\Delta T/T$ due to the Sachs-Wolfe
effect \cite{sw}:
\begin{equation}
\big\langle\left(\frac{\Delta T}{T}\right)^2\big\rangle_{10^0} = {\rm S} + 
{\rm T} \simeq 1.1 \times 10^{-10},\;\;\;\; {\rm S} = 0.04 \; \langle q^2
\rangle_{10^0}\simeq 0.06 q_{k_2}^2.
\end{equation}
The relation between  the variance of the $q$ potential averaged in 
$10^0$-angular-scale and the power spectrum at COBE scale, involves a
factor of the effective 
interval of spectral wavelengths proportional to
$\;\ln \left( \frac{k_2}{k_{hor}} \right)\simeq 1.6$.
To estimate T/S, we use the following approximation formula at 
$x_2 = x_{COBE}$:
\begin{equation}
\frac{\rm{T}}{\rm{S}} \simeq -6n_T = \frac{6}{\sqrt{c^2 + x^2_2}}\left( 1-
\frac{x_2}{\sqrt{c^2 + x^2_2}}\right).
\end{equation}

Evidently, both normalizations, (6) and (7), determine essentially the 
corresponding $q_k$ amplitudes at the locations of cluster ($k_1$) and COBE 
($k_2$) scales.
Taking their ratio we get the key equation relating $c$ and $k_{cr}$
(see eqs.(3),(4),(8)):
\begin{equation}
\left( \frac{q_{k_1}}{q_{k_2}} \right)^2 \simeq
D\left( 1 +\frac{\rm T}{\rm S}\right) ,
\end{equation}

Eq.(9) has a clear physical meaning: the ratio of the S-spectral powers at
cluster and COBE scales is proportional to $\sigma_8^2$ and inversly
proportional to the fraction of the scalar mode contributing to the 
large-scale temperature anisotropy variance, S/(S+T).
It provides quite a general constraint on the fundamental 
inflation spectra in a wide set of dark matter models using only two basic
measurement (the cluster abundance and large scale $\Delta T/T$). 
The DM information is contained in the D-coefficient which can be calculated 
using the same equation (9) for a simple inflationary spectrum (preserving the 
given DM model). For CDM with $h=0.5$ we have:
\begin{equation}
D\simeq \frac{0.6\sigma_8^2}{1-3.1\Omega_b}\; ,\;\;\;\;\;
\Omega_b <0.2.
\end{equation}

The solution of eq.(9) has been obtained in the plane $x_2 - c$ 
numerically. For the whole range $0.1<D<0.5$, it can be analytically 
approximated with a precision better that 10\% as follows:
\begin{equation}
\ln^2 \left(\frac{k_0}{k_{cr}}\right)\simeq E\left(c_0-c\right)\left(c
+c_1\right)\;,\;\;\;\; 2<c\leq c_0\;.
\end{equation}
Notice there exists no solution of eq.(9) for $c>c_0$. 
We have found the following best fit coefficients $E,\;k_0[h/\rm{Mpc}]$ and 
$c_{0,1}$:
\[
E\simeq 1,\;\;\;\ln k_0\simeq 49D^2+1.3,\;\;\;
c_0\simeq 61D^2+6.2,\;\;\; c_1\simeq 44D^2+4.0.
\]
The tensor-mode-contribution is approximated similarly ($k_{cr}$ is measured in $[h/\rm{Mpc}]$):
\begin{equation}
\frac{\rm T}{\rm S}\simeq \frac{2.53-4.3D}{(\ln k_{cr}+4.65)^{2/3}}
+\frac{1}{3}\;.
\end{equation}
%

\section{Discussion}

We have presented a new inflationary model predicting a near scale-invariant
spectrum of density perturbations and large amount of CGWs. Our model is
consistent with COBE $\Delta T/T$ and cluster abundance data. The perturbation
spectra depend on one free scale-parameter, $k_{cr}$, which can be found in 
further analysis by fitting other observational data.  At the location of 
$k_{cr}$, the CDP spectrum transfers smoothly from the red ($k<k_{cr}$) to 
the blue ($k>k_{cr}$) parts.

Today we seriously discuss a nearly flat shape of the
dimensionless CDP spectrum within the scale range encompassing clusters and 
superclusters,
\begin{equation} 
\Delta_k^2\sim k^{(0.9^+_-0.2)}\;,\;\;\;\;k \in (0.04, 0.2) h\; \rm{Mpc}^{-1},
\end{equation}
(with a break towards the HZ slope on higher scales) which stays in obvious
disagreement with the sCDM prediction.  The arguments supporting 
eq.(13) came from the analysis of large-scale galaxy distribution \cite{guzzo}
and the discovery of large quasar groups \cite{kl}${}^,$ \cite{kkl}, 
a higher statistical support was brought by recent 
measurements of the galaxy cluster power spectrum \cite{tadros}${}^,$ 
\cite{retz}.

A possible explanation of eq.(13) could be a
fundamental red power spectrum established on large scales, then the
transition to the spectrum (13) at $\sim 100 \rm{Mpc}/h$ would be much
easier understood with help of a traditional modification of the transfer 
function $T(k)$ (e.g.  for mixed hot+cold dark matter). The rediness may be  
not too high, remaining in the range (0.9, 1). A way to enhance the power
spectrum at Mpc scale could be the identification of $k_{cr}$ within a
cluster scale ($k_{cr}\sim k_1$).

 Notice that one of the problems for the matter-dominated 
models is a low number of $\sigma_8$: if $\sigma_8<0.6$, then the first 
acoustic peak in $\Delta T/T$ cannot be as high as ${}^>_\sim 70\;\mu$K.
\subsection*{Acknowledgments}
The work was partly supported by German Scientific Foundation (DFG-436 RUS
113/357/0) and INTAS grant (97-1192). V.N.L. and E.V.M. are grateful to the 
Organizing Committee for the hospitality.

\end{document}